\documentclass[english]{article}
\usepackage[T1]{fontenc}
\usepackage[latin9]{inputenc}
\usepackage{geometry}
\usepackage{mathrsfs}
\usepackage{amstext}
\usepackage{amssymb}
\usepackage{graphicx}
\usepackage{setspace}
\usepackage{esint}
\usepackage{babel}
\begin{document}
\begin{doublespace}

\title{Induced Superconductivity in the Quantum Spin Hall Edge}
\end{doublespace}

\begin{singlespace}

\author{\noindent Sean Hart$^{\dagger1}$, Hechen Ren$^{\dagger1}$, Timo
Wagner$^{1}$, Philipp Leubner$^{2}$, Mathias Mühlbauer$^{2}$,\\
Christoph Brüne$^{2}$, Hartmut Buhmann$^{2}$, Laurens W. Molenkamp$^{2}$,
Amir Yacoby$^{1}$}
\end{singlespace}

\begin{singlespace}

\date{\noindent $^{1}$Department of Physics, Harvard University, Cambridge,
MA, USA \\
$^{2}$ Physikalisches Institut (EP3), Universität Würzburg, 97074
Würzburg, Germany\\
$^{\dagger}$These authors contributed equally to this work}
\end{singlespace}
\maketitle
\begin{abstract}
Topological insulators are a newly discovered phase of matter characterized
by a gapped bulk surrounded by novel conducting boundary states \cite{Hasan2010,Qi2011,Kane2005_Z2}.
Since their theoretical discovery, these materials have encouraged
intense efforts to study their properties and capabilities. Among
the most striking results of this activity are proposals to engineer
a new variety of superconductor at the surfaces of topological insulators
\cite{Fu2008,Fu2009}. These topological superconductors would be
capable of supporting localized Majorana fermions, particles whose
braiding properties have been proposed as the basis of a fault-tolerant
quantum computer \cite{Nayak2008}. Despite the clear theoretical
motivation, a conclusive realization of topological superconductivity
remains an outstanding experimental goal. Here we present measurements
of superconductivity induced in two-dimensional HgTe/HgCdTe quantum
wells, a material which becomes a quantum spin Hall insulator when
the well width exceeds $d_{C}=6.3$ nm \cite{Konig2008}. In wells
that are 7.5 nm wide, we find that supercurrents are confined to the
one-dimensional sample edges as the bulk density is depleted. However,
when the well width is decreased to 4.5 nm the edge supercurrents
cannot be distinguished from those in the bulk. These results provide
evidence for superconductivity induced in the helical edges of the
quantum spin Hall effect, a promising step toward the demonstration
of one-dimensional topological superconductivity. Our results also
provide a direct measurement of the widths of these edge channels,
which range from 180 nm to 408 nm.
\end{abstract}
Topological superconductors, like topological insulators, possess
a bulk energy gap and gapless surface states. In a topological superconductor,
the surface states are predicted to manifest as zero-energy Majorana
fermions, fractionalized modes which pair to form conventional fermions.
Due to their non-Abelian braiding statistics, achieving control of
these Majorana modes is desirable both fundamentally and for applications
to quantum information processing. Proposals toward realizing Majorana
fermions have focused on their emergence within fractional quantum
Hall states \cite{Moore1991} and spinless $p+ip$ superconductors
\cite{Read2000}, and on their direct engineering using $s$-wave
superconductors combined with topological insulators or semiconductors
\cite{Sau2010,Alicea2010}. Particularly appealing are implementations
in one-dimensional (1D) systems, where Majorana modes would be localized
to the ends of a wire. In such a 1D system, restriction to a single
spin degree of freedom combined with proximity to an $s$-wave superconductor
would provide the basis for topological superconductivity \cite{Kitaev2001}.
Effort in this direction has been advanced by studies of nanowire
systems \cite{Mourik2012,Das2012,Rokhinson2012,Lee2012,Churchill2013,Finck2013}
and by excess current measurements on InAs/GaSb devices \cite{Knez2012}.
Given the wide interest in Majorana fermions in one dimension, it
is essential to expand the search to other systems whose properties
are suited toward their control.

An attractive route toward a 1D topological superconductor uses as
its starting point the two-dimensional (2D) quantum spin Hall (QSH)
insulator. This topological phase of matter was recently predicted
\cite{Bernevig2006_PRL,Bernevig2006_Science} and observed \cite{Konig2007,Roth2009}
in HgTe/HgCdTe quantum wells thicker than a critical thickness $d_{C}=6.3$
nm. Due to strong spin-orbit coupling the bulk bands of the system
invert, crossing only at the edges of the system to form 1D counterpropagating
helical modes. Time-reversal symmetry ensures protection of these
modes against elastic backscattering over distances shorter than the
coherence length \cite{Schmidt2012}. The helical nature of the edge
modes makes them a particularly appealing path toward the realization
of a topological superconductor, due to the intrinsic elimination
of their spin degree of freedom. Here we report measurements of supercurrents
confined to edge states in HgTe/HgCdTe quantum well heterostructures,
a critical step toward the demonstration of 1D topological superconductivity.

Our approach consists of a two-terminal Josephson junction, with a
rectangular section of quantum well located between two superconducting
leads (Figure 1). At a given bulk carrier density, the presence or
absence of helical edge channels influences the supercurrent density
profile across the width of the junction. In the simplest case the
supercurrent density is uniform throughout the device, and edge channels
are indistinguishable from bulk channels (Figure 1a). This behavior
would be expected for a non-topological junction (quantum well width
smaller than $d_{C}$), or in a topological junction (quantum well
width larger than $d_{C}$) far from the bulk insulating regime.

In a topological junction, decreasing the bulk carrier density brings
the device closer to the QSH insulator regime (Figure 1b). Scanning
SQUID measurements suggest that over a range of bulk densities the
QSH edge channels coexist with bulk states, and can carry considerably
more edge current than would be expected for a non-topological conductor
\cite{Nowack2013}. In the two-terminal configuration, this helical
edge contribution appears as peaks in the supercurrent density at
each edge. When the bulk density becomes sufficiently low, these edge
peaks are the only features in the supercurrent density (Figure 1c).
Then the supercurrent is carried solely along the helical edges, and
the system is in the regime of the quantum spin Hall superconductor.

Placing such a Josephson junction in a perpendicular magnetic field
$B$ provides a way to measure the supercurrent density in the quantum
well. In general, the maximum supercurrent that can flow through a
Josephson junction is periodically modulated by a magnetic field.
Typically, the period of the modulation corresponds to the magnetic
flux quantum $\Phi_{0}=h/2e$. In our junctions this period matches
the area of the HgTe region plus half the area occupied by each contact,
a result of the Meissner effect. The particular shape of the critical
current interference pattern depends on the phase-sensitive summation
of the supercurrents traversing the junction \cite{Tinkham2004}.
In the case of a symmetric supercurrent distribution, this integral
takes the simple form:

\begin{center}
$I_{C}^{max}(B)=\left|\int_{-\infty}^{\infty}dxJ_{S}(x)\cos(2\pi L_{J}Bx/\Phi_{0})\right|.$
\par\end{center}

\noindent Here $L_{J}$ is the length of the junction along the direction
of current, accounting for the magnetic flux focusing from the contacts. 

It is evident that different supercurrent densities $J_{S}(x)$ in
the junction can give rise to different interference patterns $I_{C}^{max}(B)$.
The flat supercurrent density of a trivial conductor corresponds to
a single-slit Fraunhofer pattern $\left|(\sin(\pi L_{J}BW/\Phi_{0}))/(\pi L_{J}BW/\Phi_{0})\right|$,
characterized by a central lobe width of $2\Phi_{0}$ and side lobes
decaying with $1/B$ dependence (Figure 1a). As helical edges emerge,
this single-slit interference evolves toward the more sinusoidal oscillation
characteristic of a SQUID (Figure 1b). The central lobe width shrinks
to $\Phi_{0}$ when only edge supercurrents remain, with the side
lobe decay determined by the widths of the edge channels (Figure 1c).
Measuring the dependence of $I_{C}^{max}$ on $B$ therefore provides
a convenient way to measure the distribution of supercurrent in a
junction. To quantitatively extract $J_{S}(x)$ from the measured
quantity $I_{C}^{max}(B)$ we follow an approach developed by Dynes
and Fulton, where nonzero $I_{C}^{max}(B)$ minima are ascribed to
an asymmetric supercurrent distribution \cite{Dynes1971}. Although
other effects may lead to nonzero minima in $I_{C}^{max}(B)$, we
consider here only the possibility of an odd component in $J_{S}(x)$.
Full details of the extraction procedure can be found in the Supplementary
Information.

To study how supercurrents flow in the QSH regime, we measure a Josephson
junction consisting of a 7.5 nm-wide quantum well contacted by titanium/aluminum
leads. Our contact lengths are each 1 $\mu$m, and the contact separation
is 800 nm. The junction width of 4 $\mu$m is defined by etched mesa
edges. A voltage $V_{G}$ applied to a global topgate allows us to
tune the carrier density in the junction. At each value of $V_{G}$
and $B$, the critical current $I_{C}^{max}$ is determined by increasing
the current through the junction while monitoring the voltage across
the leads. The behavior observed in this device is reproducable in
several other similar junctions, as reported in the Supplementary
Information. 

As a function of the topgate voltage, the overall behavior of the
junction evolves between two extremes. At more positive gate voltage
and higher bulk density, the critical current envelope strongly resembles
a single-slit pattern (Figure 2a). This type of interference suggests
a nearly uniform supercurrent density throughout the sample, confirmed
by transformation to the $J_{S}(x)$ picture (Figure 2b). This nearly
flat distribution indicates that the quantum well is in the high carrier
density regime of an essentially trivial conductor.

At more negative gate voltage and lower bulk density, the critical
current envelope becomes close to a sinusoidal oscillation (Figure
2c). The shift toward a SQUID-like interference pattern corresponds
to the development of sharp peaks in supercurrent density at the mesa
edges (Figure 2d).

We can track this evolution in a single device by measuring the critical
current envelope at a series of gate voltages. As the topgate is varied
from $V_{G}=1.05$ V to $V_{G}=-0.45$ V, the maximum critical current
decreases from 505 nA to 5.7 nA. At the same time, the overall critical
current behavior shows a narrowing of the central interference lobe,
from $2\Phi_{0}$ at positive gate voltages to $\Phi_{0}$ at negative
gate voltages (Figure 3a,b). The side lobes additionally become continuously
more pronounced, indicating the confinement of supercurrent to channels
at the edges of the junction (Figure 3c,d). The normal resistance,
measured at large bias to overcome superconductivity, increases from
160 $\Omega$ to $\sim3,000$ $\Omega$ over the range of this transition.
While it is possible to gate further toward depletion, the critical
currents become too small to reliably measure and no meaningful supercurrent
density can be extracted.

At the most negative gate voltage, $V_{G}=-0.45$ V, we can estimate
the widths of the supercurrent-carrying edge channels using a Gaussian
lineshape (Figure 3f). Using this method, we find widths of 408 nm
and 319 nm for the two edges. Our measurements of edge widths in another
device with similar dimensions, as well as one with a 2 $\mu$m mesa
width, show edges as narrow as 180 nm (see Supplementary Information).
These width variations, as well as the normal state resistance that
is low compared to the resistance $h/2e^{2}$ for two ballistic 1D
channels, suggest the presence of additional edge modes or of bulk
modes coupled too weakly across the junction to carry supercurrent.

To provide further evidence that the observed edge supercurrents are
topological in nature, we next turn to a heterostructure with a quantum
well width of 4.5 nm. In this device, the well width is smaller than
the critical width $d_{C}$, so that the sample is not expected to
enter the QSH regime. Near zero topgate voltage and a normal resistance
of 270 $\Omega$, the critical current interference pattern has a
maximum of 243 nA and resembles a single-slit envelope (4a,b). Upon
energizing the topgate and decreasing the bulk density, the single-slit
pattern persists. In contrast to the wide well sample, this behavior
corresponds to a supercurrent density that remains distributed throughout
the junction even as the normal resistance rises to several k$\Omega$
(Figure 4c-f). Because the edge supercurrents are present only when
the well width is larger than $d_{C}$, we conclude that our observations
provide evidence for induced superconductivity in the helical QSH
edge states.

The ability to induce supercurrents through the helical edges of a
two-dimensional topological insulator represents a significant step
toward the realization of topological superconductivity. A one-dimensional
system with superconducting pairing and only one spin degree of freedom
should be capable of entering a topological phase, allowing Majorana
zero-modes at its ends. The HgTe/HgCdTe system represents a natural
host for these effects, since its edge modes occur in the same manner
as the paired electrons of an $s$-wave superconductor. Our observation
of supercurrents confined to the edges of topologically nontrivial
HgTe quantum wells distinguishes this system as an especially promising
platform in which to study the physics emerging from interactions
among electrons in reduced dimensions.

\bibliographystyle{naturemag}
\bibliography{qsh_supcurrent_oct30_2013}

\begin{doublespace}
\textbf{Acknowledgments: }We acknowledge Anton Akhmerov and Jay Deep
Sau for theoretical discussions. We acknowledge financial support
from Microsoft Corporation Project Q, the NSF DMR-1206016, the DOE
SCGF Program, the German Research Foundation (DFG-JST joint research
program ``Topological Electronics''), and EU ERC-AG program (project
3-TOP).

\textbf{Author Contributions: }All authors contributed collaboratively
to the work.

\textbf{Author Information: }The authors declare no competing financial
interests. Correspondence and requests for materials should be addressed
to yacoby@physics.harvard.edu.

\begin{figure}[p]
\centering{}\includegraphics[scale=0.15]{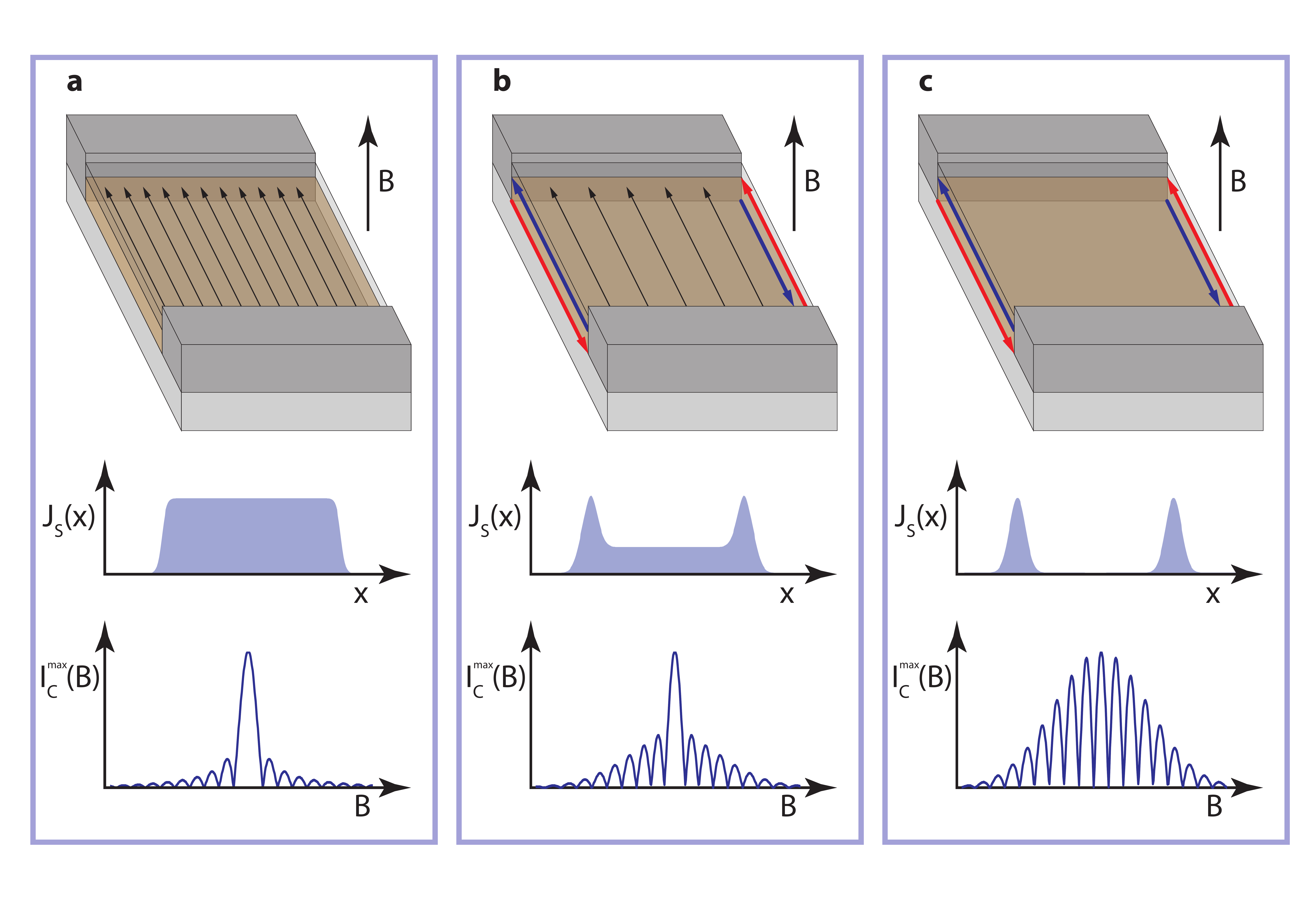}\caption{Expected two-terminal behavior in different regimes of a topological
quantum well. a, When the bulk of a sample is filled with charge carriers,
supercurrent can flow uniformly across the junction, corresponding
to a flat supercurrent density $J_{S}(x)$. A perpendicular magnetic
field $B$ modulates the maximum critical current $I_{C}^{max}$,
resulting in a single-slit Fraunhofer interference pattern. b, As
the bulk carriers are depleted, the supercurrent density develops
peaks due to the presence of the helical edges. This evolution toward
edge-dominated transport appears in the interference pattern as a
narrowing central lobe width and more pronounced side lobe amplitudes.
c, When no bulk carriers remain, the supercurrent is carried only
along the helical edges. In this regime the interference results in
a sinusoidal double-slit pattern, with an overall decay in $B$ that
is determined by the width of the edge channels.}
\end{figure}

\end{doublespace}

\begin{figure}[p]
\begin{centering}
\includegraphics[scale=0.5]{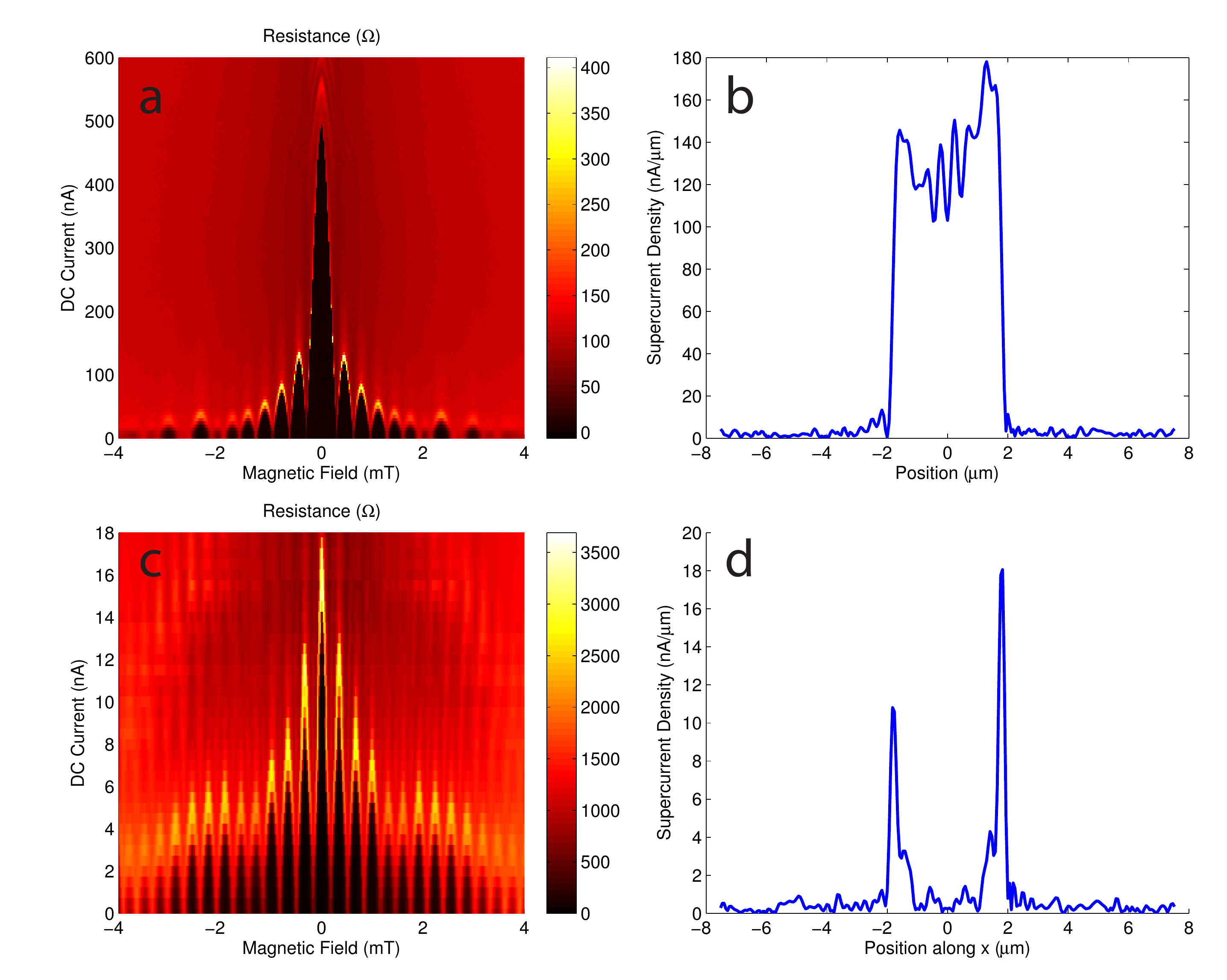}
\par\end{centering}

\caption{General behavior observed in the topological Josephson junction. a,
A map of the differential resistance across the junction, measured
with the topgate at $V_{G}=1.05$ V, shows the single-slit interference
characteristic of a uniform supercurrent density. b, The supercurrent
density, extracted for $V_{G}=1.05$ V, is consistent with trivial
charge transport throughout the bulk of the junction. c, When the
topgate voltage is lowered to $V_{G}=-0.425$ V, the differential
resistance shows a more sinusoidal interference pattern. d, Using
the inteference envelope measured at $V_{G}=-0.425$ V, the supercurrent
density is clearly dominated by the contribution from the edges. In
this regime almost no supercurrent passes through the bulk.}
\end{figure}

\begin{figure}[p]
\begin{centering}
\includegraphics[scale=0.4]{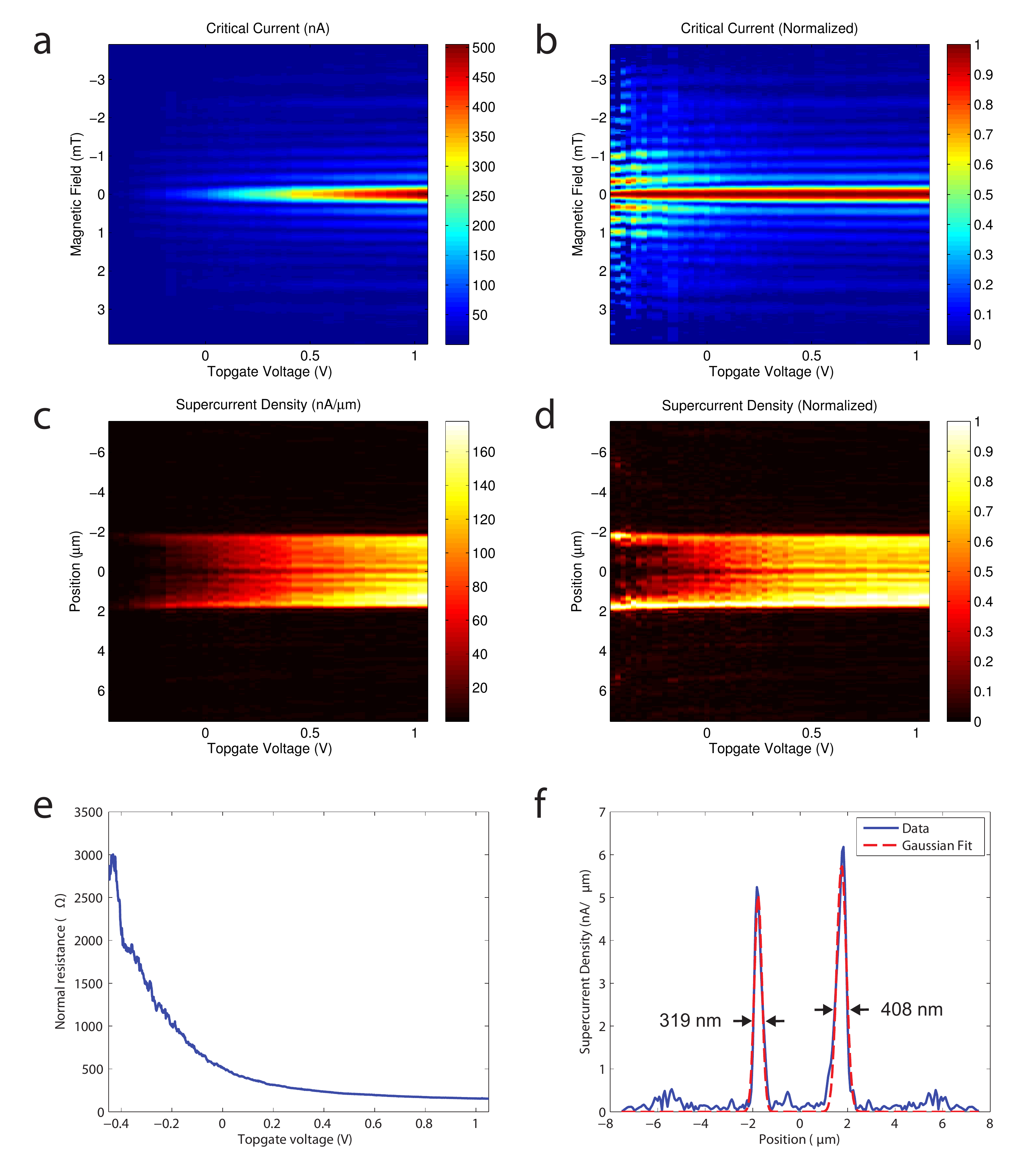}
\par\end{centering}

\caption{Continuous evolution with gating in the topological Josephson junction.
a, As the topgate is varied from $V_{G}=1.05$ V to $V_{G}=-0.45$
V, the maximum critical current decreases from 505 nA to 5.7 nA. b,
Normalizing the interference patterns to their peak values reveals
the evolution toward sinusoidal interference. c, Using the envelope
at each gate voltage, the evolution of the supercurrent density can
be visualized. d, By normalizing each supercurrent density to its
maximum value, the transition from trivial to edge-dominated supercurrent
transport can be clearly seen. e, This transition occurs as the normal
device resistance increases from 160 $\Omega$ to 3,000 $\Omega$.
f, At the most negative gate voltage, $V_{G}=-0.45$ V, the supercurrent
density provides a measurement of the edge widths.}
\end{figure}

\begin{figure}[p]

\centering{}\includegraphics[scale=0.4]{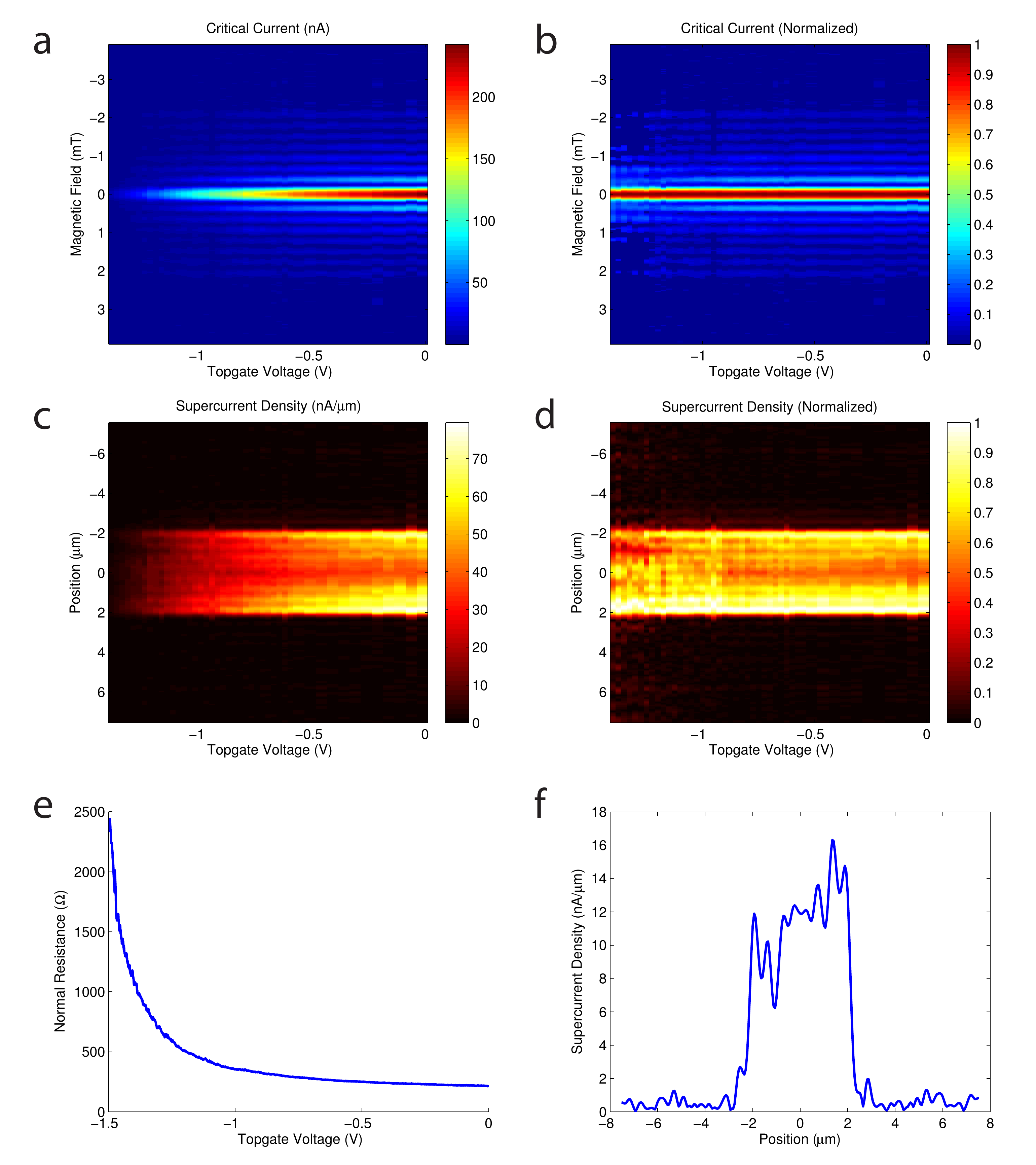}\caption{Continuous evolution with gating in the non-topological Josephson
junction. a, As the topgate is varied from $V_{G}=0$ V to $V_{G}=-1.5$
V, the maximum critical current decreases from 243 nA to 4.4 nA. b,
Normalizing the interference patterns to their peak values shows the
stability of the single-slit pattern over a wide range of gating.
c, Using the envelope at each gate voltage, the evolution of the supercurrent
density can be visualized. d, Normalizing each supercurrent density
to its maximum value shows that the supercurrent remains distributed
throughout the device. e, This roughly uniform supercurrent distribution
remains even as the device resistance increases from 215 $\Omega$
to almost 2,500 $\Omega$. f, A linetrace of the supercurrent density
close to depletion further demonstrates that the supercurrent flows
throughout the device.}
\end{figure}

\newpage{}

\begin{doublespace}
\begin{center}
{\LARGE{Supplementary Information for Induced Superconductivity in
the Quantum Spin Hall Edge}}
\par\end{center}{\LARGE \par}
\end{doublespace}

\begin{center}
{\large{Sean Hart$^{\dagger1}$, Hechen Ren$^{\dagger1}$, Timo Wagner$^{1}$,
Philipp Leubner$^{2}$, Mathias Mühlbauer$^{2}$,}}\\
{\large{Christoph Brüne$^{2}$, Hartmut Buhmann$^{2}$, Laurens W.
Molenkamp$^{2}$, Amir Yacoby$^{1}$}}
\par\end{center}{\large \par}

\begin{center}
{\large{$^{1}$Department of Physics, Harvard University, Cambridge,
MA, USA }}\\
{\large{$^{2}$ Physikalisches Institut (EP3), Universität Würzburg,
97074 Würzburg, Germany}}\\
{\large{$^{\dagger}$These authors contributed equally to this work}}
\par\end{center}{\large \par}

\section*{Device characteristics}

Devices were fabricated on two different HgTe/HgCdTe heterostructures,
with layer structures shown in Supplementary Figure 1. Wafer I contained
a 7.5 nm quantum well with an electron density of $3.6\times10^{11}/\text{cm}^{2}$
and a mobility of $300,000\text{ cm}^{2}/\text{Vs}$. Wafer II contained
a 4.5 nm quantum well with an electron density of $3.5\times10^{11}/\text{cm}^{2}$
and a mobility of $100,000\text{ cm}^{2}/\text{Vs}$.

Device processing consisted of the following steps. Mesas were defined
by etching with an Ar ion source, and were 100 nm in height. Contacts
consisted of 10 nm of titanium under 180 nm of aluminum, deposited
by thermal evaporation after in situ cleaning with an Ar ion source.
A 50 nm layer of aluminum oxide deposited by atomic layer deposition
isolated the mesa and contacts from the topgate, which consisted of
10 nm of titanium under 250 nm of gold. An SEM image of a junction
is depicted in Supplementary Figure 2.

\section*{Critical current measurement}

Measurements were performed in a dilution refrigerator with a base
temperature of 10 mK, and an electron temperature of 20 mK measured
using standard Coulomb blockade techniques. At each voltage $V_{G}$
on the topgate, the magnetic field was stepped through $B=0$ mT over
a range of 8 mT. At each value of magnetic field, the DC current $I_{DC}$
through a junction was then increased while monitoring the DC voltage
drop $V_{DC}$ across the junction. A voltage threshold was used to
determine the critical current; the point beyond which $V_{DC}$ was
increasing and above the threshold voltage was recorded as the critical
current $I_{C}^{max}(B,V_{G})$. Our threshold was set at 1 $\mu$V,
several standard deviations above the noise level. There is an artificial
offset introduced by this method when the critical current falls to
zero. In our analysis these artificial offset currents are reported
as zero instead of the value given by the threshold method.

\section*{Analysis of current density profile}

In a Josephson junction immersed in a perpendicular magnetic field
$B$, the magnitude of the maximum critical current $I_{C}^{max}(B)$
depends strongly on the supercurrent density between the leads. For
example, a uniform supercurrent density generates single-slit Fraunhofer
interference, while a sinusoidal double-slit interference pattern
arises from two supercurrent channels enclosing the junction area.
In the following discussion we elaborate on this correspondence, outlining
the quantitative way in which we convert our measured interference
patterns to their originating supercurrent density profiles. We assume
throughout that the current density varies only along the $x$ direction,
and that the supercurrent is directed along the orthogonal $y$ direction.
The junction then has a length $L$ in the $y$ direction, and the
leads each have a length $L_{Al}$. Our method follows the approach
developed by Dynes and Fulton \cite{Dynes1971}.

At a fixed magnetic field, the total critical current through the
Josephson junction is a phase-sensitive summation of supercurrent
over the width of the junction. Suppose we have a supercurrent density
profile $J_{S}(x)$. Then its complex Fourier transform yields a complex
critical current function $\mathscr{I}_{C}(\beta)$,

\begin{equation}
\mathscr{I}_{C}\mathrm{(\beta)}=\int_{-\infty}^{\infty}dxJ_{S}(x)e^{i\beta x},
\end{equation}

where the normalized magnetic field unit $\beta=2\pi(L+L_{Al})B/\Phi_{0}$,
and the magnetic flux quantum $\Phi_{0}=h/2e$. 

The experimentally observed $I_{C}^{max}(\beta)$ is the magnitude
of this summation: $I_{C}^{max}(\beta)\mathscr{=\mathrm{|}I}_{C}(\beta)|.$
Therefore to extract the supercurrent density from $I_{C}^{max}(\beta)$
it is necessary to first recover the complex critical current $\mathscr{I}_{C}(\beta)$. 

This reduces to a particularly simple problem in the case of an even
current density, $J_{E}(x),$ representing a symmetric distribution.
The odd part of $e^{i\beta x}$ vanishes from the integral, and equation
(1) becomes $\mathscr{I}_{C}(\beta)=I_{E}=\int_{-\infty}^{\infty}dxJ_{E}(x)\cos\beta x$.
Since $J_{E}(x)$ is real and positive, we see that $\mathscr{I}_{C}(\beta)$
is also real, and it typically alternates between positive and negative
values at each zero-crossing. Because $I_{C}^{max}(\beta)=|\mathscr{I}_{C}(\beta)|,$
we can therefore recover the exact $\mathscr{I}_{C}(\beta)$ by flipping
the sign of every other lobe of the observed $I_{C}^{max}(\beta).$

Now suppose that on top of this even function, the current distribution
has a small but non-vanishing odd component, $J_{O}(x),$ with its
Fourier transform $I_{O}(\beta)=\int_{-\infty}^{\infty}dxJ_{O}(x)\sin\beta x.$
Then (1) gives

\begin{equation}
\mathscr{I}_{C}(\beta)=I_{E}(\beta)+iI_{O}(\beta).
\end{equation}

The observed critical current $I_{C}^{max}(\beta)=\sqrt{I_{E}^{2}(\beta)+I_{O}^{2}(\beta)}$
is therefore dominated by $I_{E}(\beta)$ except at its minima points.
Approximately, $I_{E}(\beta)$ is obtained by multiplying $I_{C}^{max}(\beta)$
with a flipping function that switches sign between adjacent lobes
of the envelope function (Supplementary Figures 3,4a). When $I_{E}(\beta)$
is minimal, the odd part $I_{O}(\beta)$ dominates the critical current.
$I_{O}(\beta)$ can then be approximated by interpolating between
the minima of $I_{C}^{max}(\beta)$, and flipping sign between lobes
(Supplementary Figure 4b). A Fourier transform of the resulting complex
$\mathscr{I}_{C}(\beta)$, over the sampling range $b$ of $\beta$,
yields the current density profile (Supplementary Figure 5):

\begin{equation}
J_{S}(x)=\left|\frac{1}{2\pi}\int_{-b/2}^{b/2}d\beta\mathscr{I}_{C}\mathrm{(\beta)}e^{-i\beta x}\right|.
\end{equation}

\section*{Gating of resistance and supercurrent}

To study the variation of the normal resistance as a function of the
bulk carrier density, we swept the topgate voltage in the topological
junction (main text) from $V_{G}=1.05$ V to $V_{G}=-3$ V. Over this
gate range, the differential resistance was measured using an AC excitation
of 5 nA. We additionally maintained a constant DC voltage bias of
750 $\mu$V across the junction to avoid features related to superconductivity.
The resulting normal resistance measurement displays two relatively
conductive regimes separated by a resistance plateau peaking near
6-8 k$\Omega$ (Supplementary Figure 6a). This behavior is consistent
with previous transport measurements of the QSH effect, where the
QSH insulator state appears as a resistance peak when samples are
gated from n-type to p-type regimes \cite{Konig2007}. The value of
the resistance plateau is lower than the expected resistance $h/2e^{2}$
for two ballistic 1D channels, suggesting that additional bulk modes
are present. Near $V_{G}=-3$ V, our junction resistance saturates
at 3 k$\Omega$ and we observe no superconductivity. This behavior
can be explained by the formation of an n-p-n junction, where barriers
between regions of different carrier type can block the transmission
of supercurrent.

As we tune the topgate to more negative voltages, the maximum critical
current of our junction decreases (Figure 3a). The electron temperature
$T=20$ mK provides an estimate $2ek_{B}T/\hbar\approx$1 nA for the
smallest critical currents that can still be reliably measured. For
the topological junction shown in the main text (Figures 2, 3), this
limit is reached above a topgate voltage of $V_{G}=-0.45$ V. However,
even beyond this point clear magnetoresistance oscillations are still
apparent. In Supplementary Figure 6b these oscillations are plotted
for $V_{G}=-0.7$ V. The magnetic field period corresponds to the
magnetic flux quantum $\Phi_{0}=h/2e$ observed throughout the gating
of the device, suggesting that supercurrent transport persists well
into the QSH regime.

\section*{Additional devices}

In addition to the two devices presented in the main text, we also
measured several different junction geometries fabricated using the
7.5 nm quantum well heterostructure. One of these junctions had a
width of 2 microns, but was otherwise identical to the topological
device presented in the main text. This device also showed a transition
from uniform bulk supercurrent to edge-dominated supercurrent, concurrently
with the normal resistance rising from 300 ohms to 4,000 ohms (Supplementary
Figure 7). The size of the magnetic field period in this device is
0.68 mT, consistent with the overall device area of 2 microns $\times$
(800 nm + 1 micron). From the supercurrent density profile in the
QSH regime, we extract edge widths of 180 nm and 197 nm.

The other device, a 4 micron wide junction, was also identical to
the topological junction from the main text except that the topgate
was only 200 nm long and was threaded between the contacts. Although
this topgate did not fully cover the junction, the behavior observed
in this device was still consistent with the other topological devices
(Supplementary Figure 8). This suggests that the gate effect was approximately
uniform across the area between the contacts. When the normal resistance
of the device was 4,000 ohms, supercurrent transport was observed
in this device through edges with widths of 208 nm and 214 nm. Even
after supercurrents became too small to measure, the normal resistance
of this device approached the expected value of $h/2e^{2}$ for transport
through two ballistic one-dimensional edge modes.

\pagebreak{}

\date{\noindent 
\begin{figure}[p]
\begin{centering}
\includegraphics[scale=0.5]{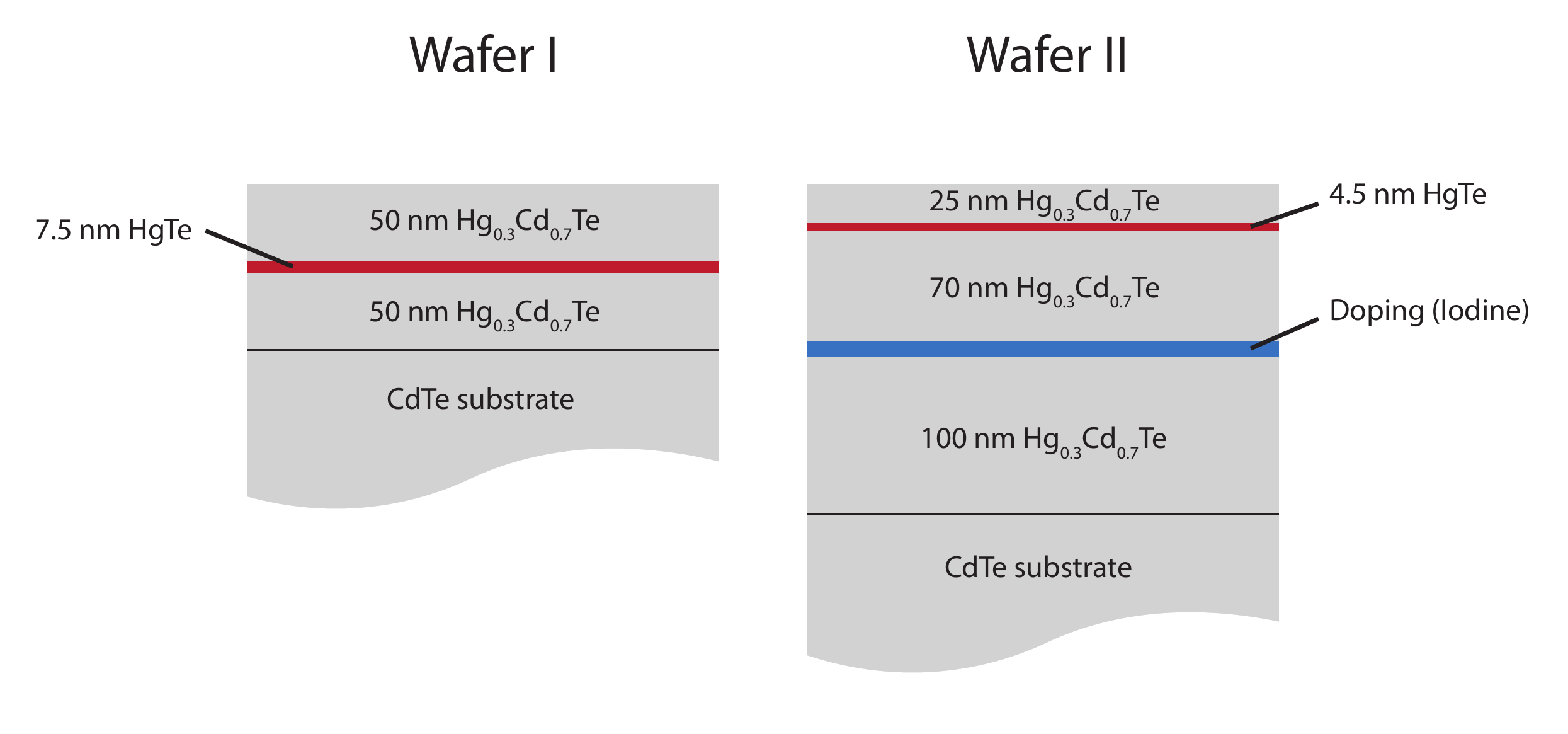}
\par\end{centering}

Supplementary Figure 1: Schematics of the heterostructures used in
the experiment. The quantum well thicknesses were 7.5 nm for Wafer
I and 4.5 nm for Wafer II.
\end{figure}
\begin{figure}[p]
\begin{centering}
\includegraphics[scale=0.5]{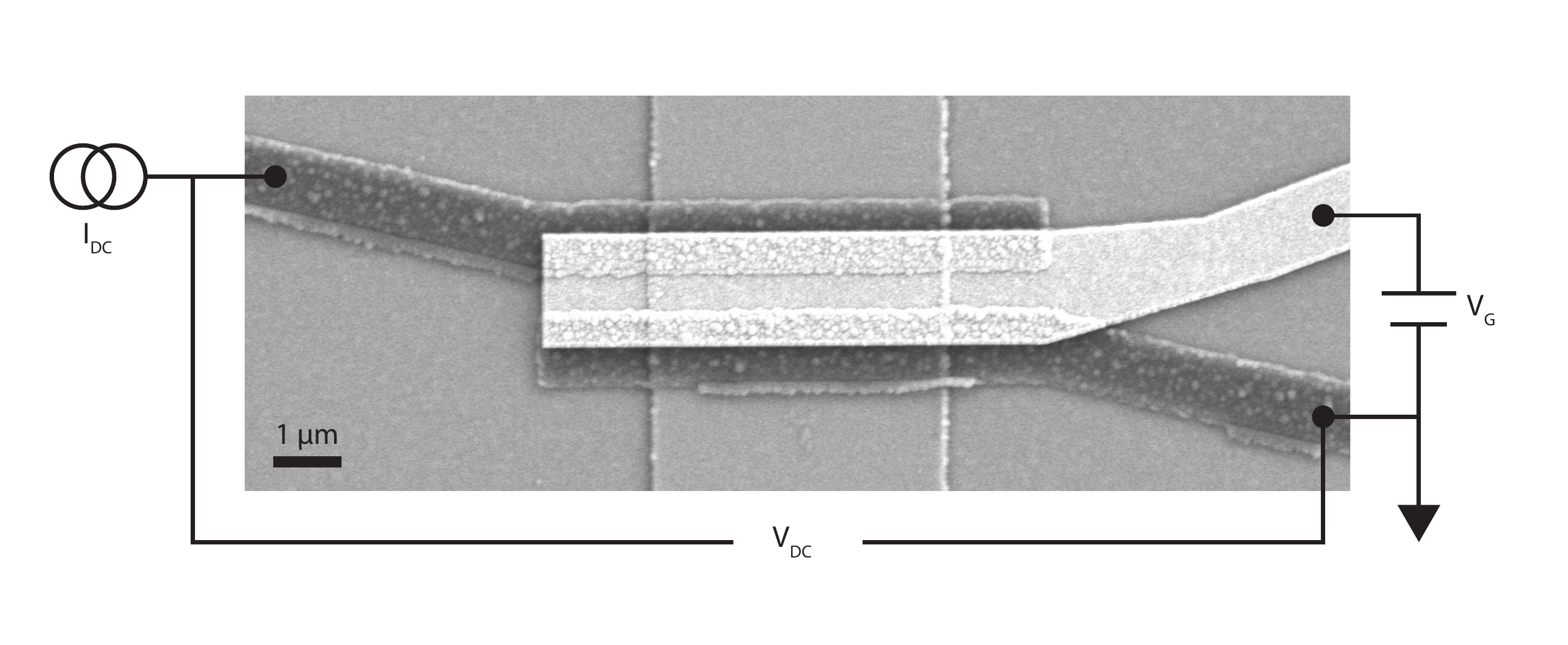}
\par\end{centering}

Supplementary Figure 2: A scanning electron micrograph showing the
layout of the junctions. A mesa 4 microns in width was contacted by
Ti/Al leads. The voltage drop $V_{DC}$ across these leads was monitored
as a function of the DC current $I_{DC}$ flowing between them. A
voltage $V_{G}$ applied to a topgate was used to tune the electron
density in the device. 
\end{figure}
}

\date{\noindent 
\begin{figure}[p]
\begin{centering}
\includegraphics[scale=0.5]{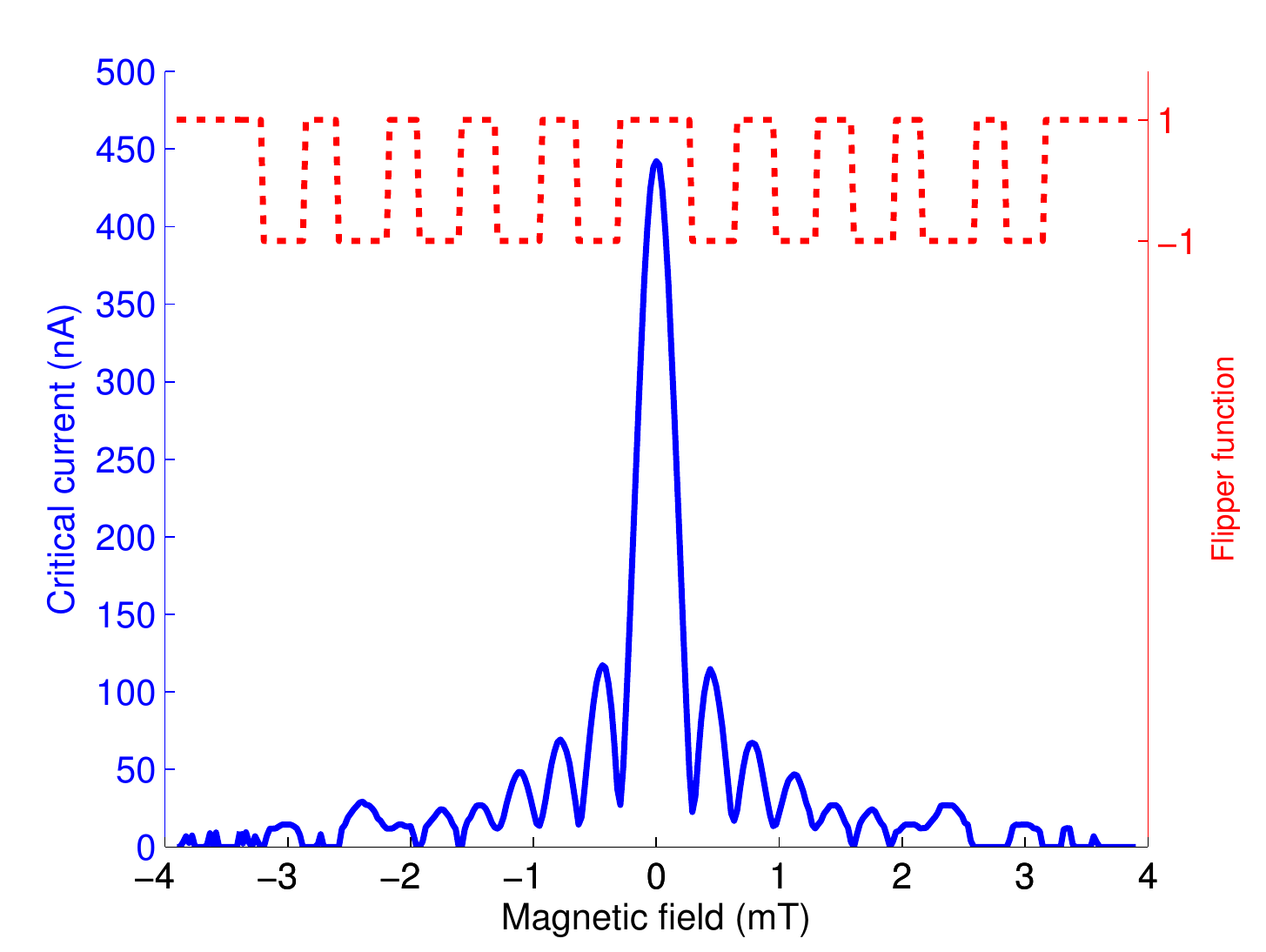}
\par\end{centering}

Supplementary Figure 3: Recovering the critical current phase. When
the current distribution is mostly symmetric, the experimentally observed
critical current envelope (blue line) approaches zero between peaks.
In such cases a flipping function (red dashed line) that changes sign
at each node of the envelope enables the recovery of $\mathscr{I}_{C}(B)$
from $I_{C}^{max}(B).$
\end{figure}
\begin{figure}[p]
\begin{centering}
\includegraphics[scale=0.4]{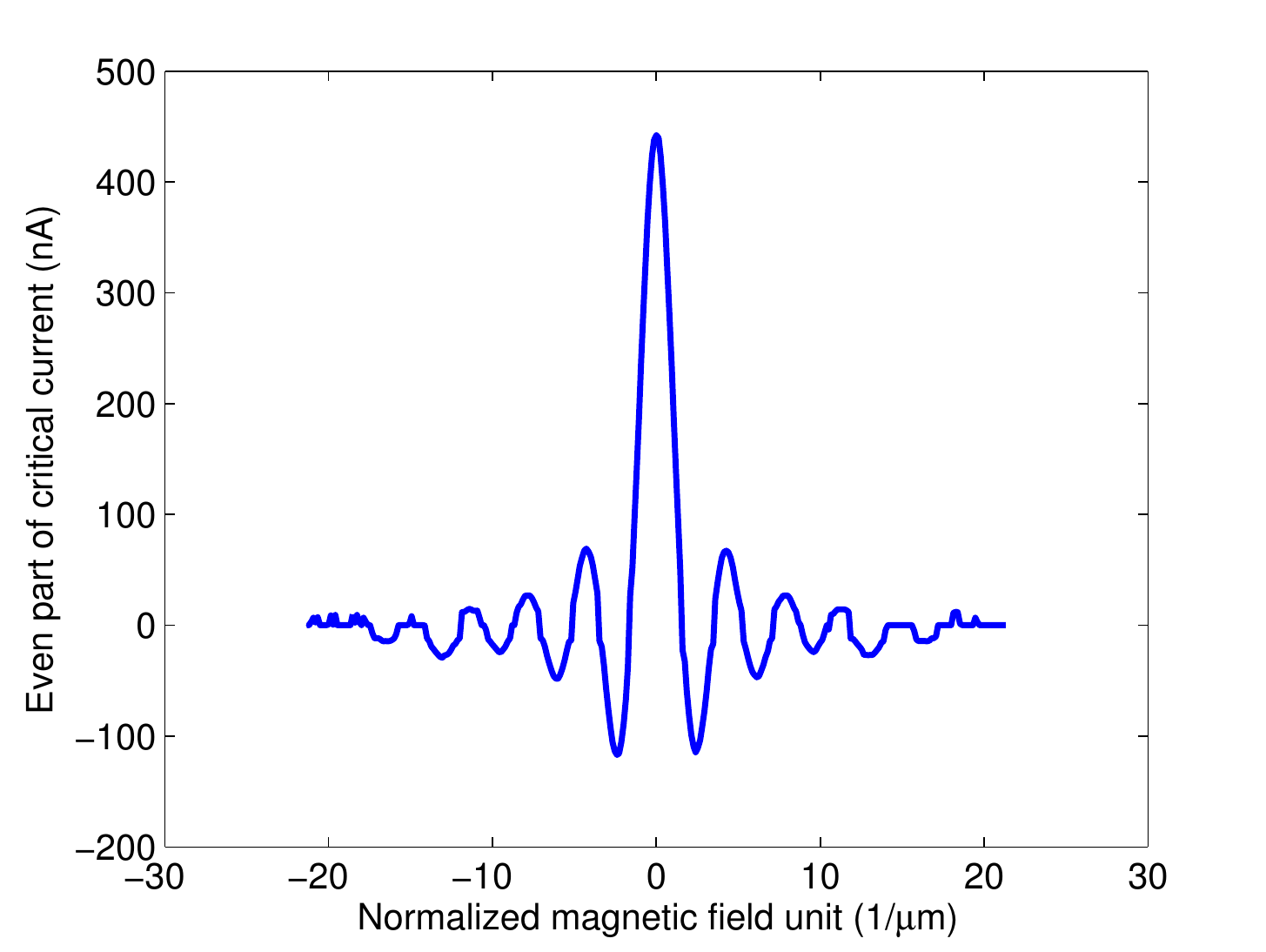}\includegraphics[scale=0.4]{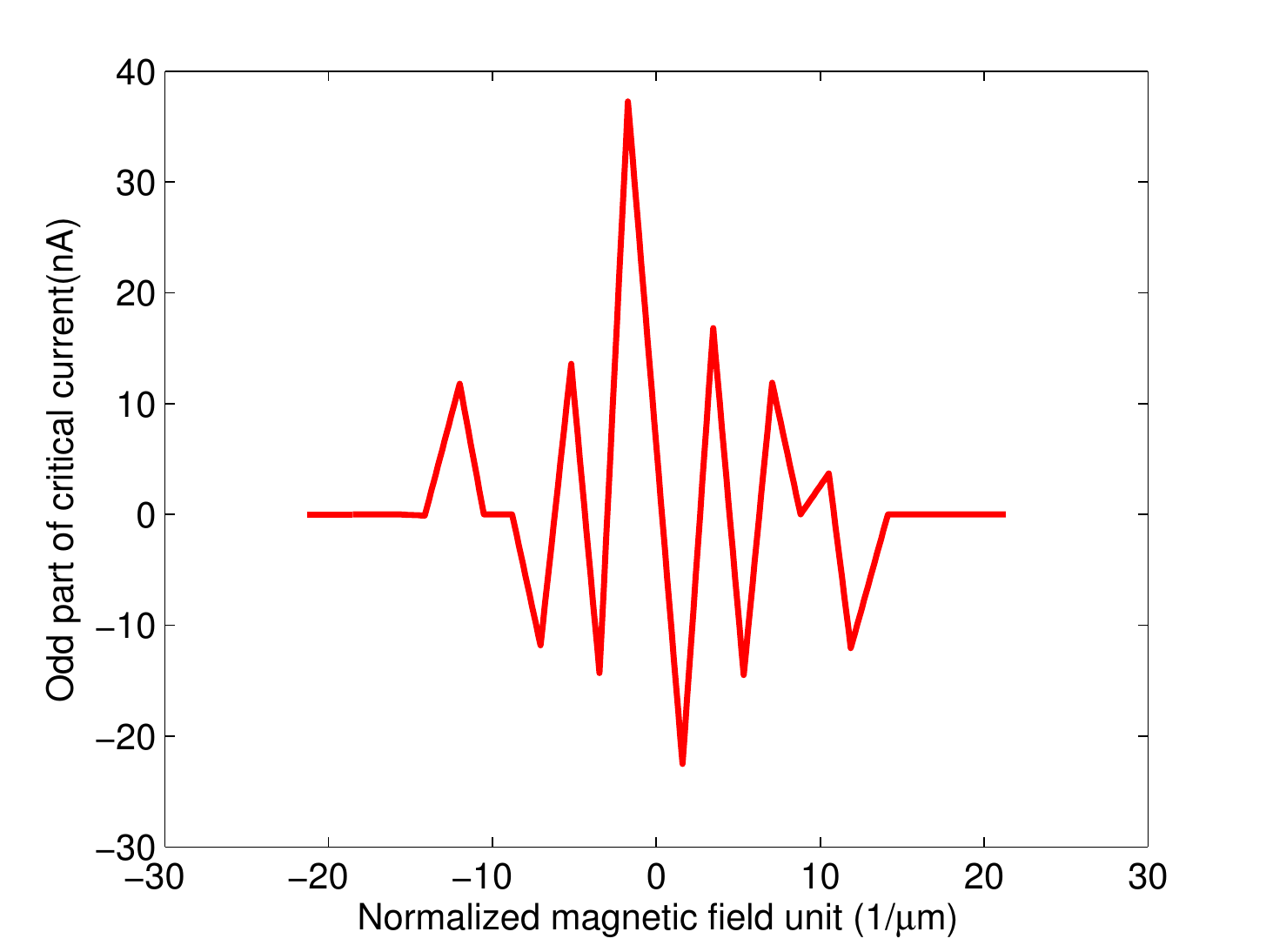}
\par\end{centering}

Supplementary Figure 4: Recovered complex critical current. a, The
recovered critical current $I_{E}(\beta)$ that corresponds to the
even part of the current density profile $J_{E}(x)$. b, The recovered
critical current $I_{O}(\beta)$ that corresponds to the odd part
of the current density profile $J_{O}(x)$.
\end{figure}
\begin{figure}[p]
\begin{centering}
\includegraphics[scale=0.5]{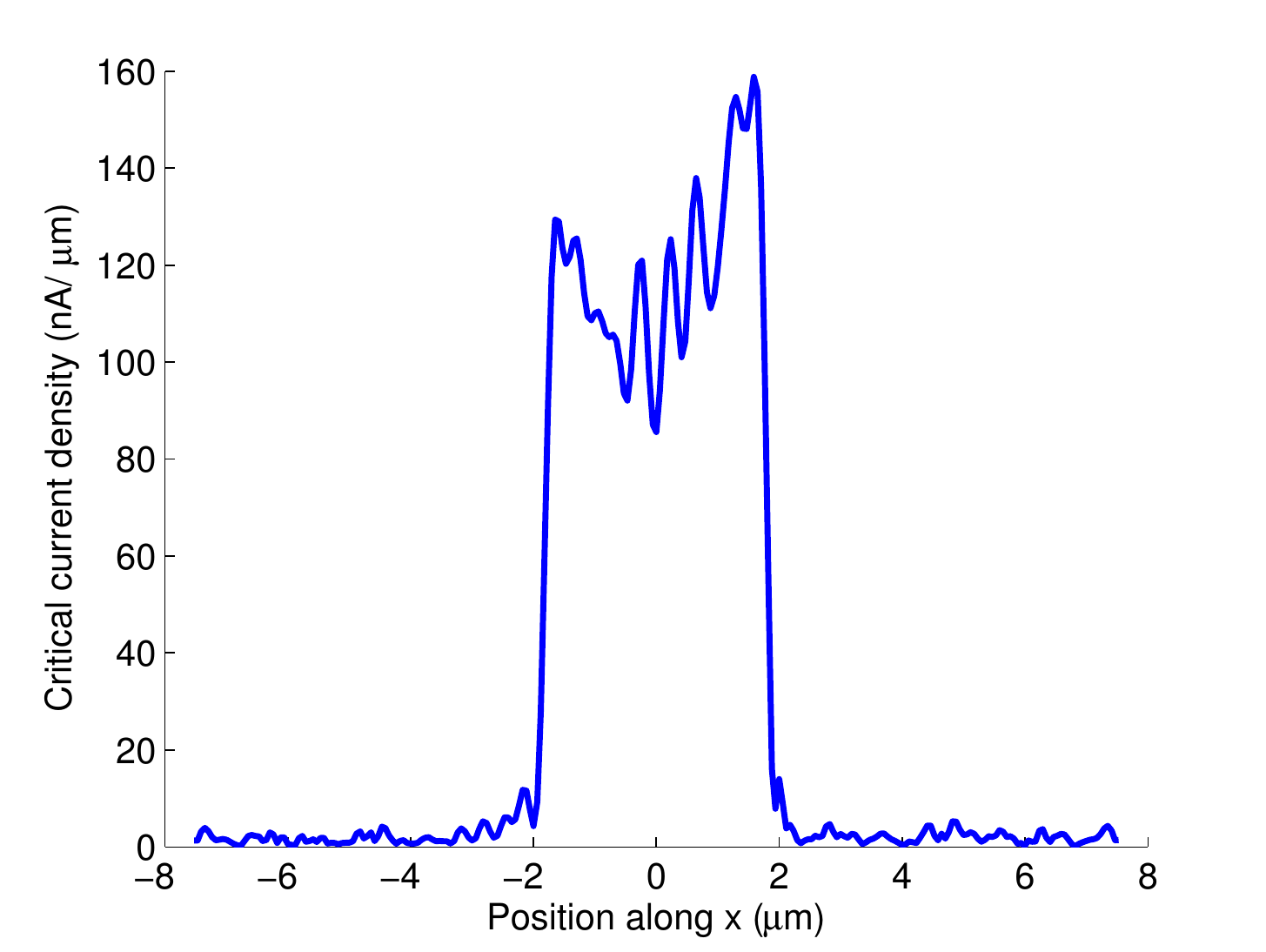}
\par\end{centering}

Supplementary Figure 5: The current density profile $J_{S}(x)$ that
corresponds to the envelope in Supplementary Figure 3.
\end{figure}
}

\date{\noindent 
\begin{figure}[p]
\begin{centering}
\includegraphics[scale=0.5]{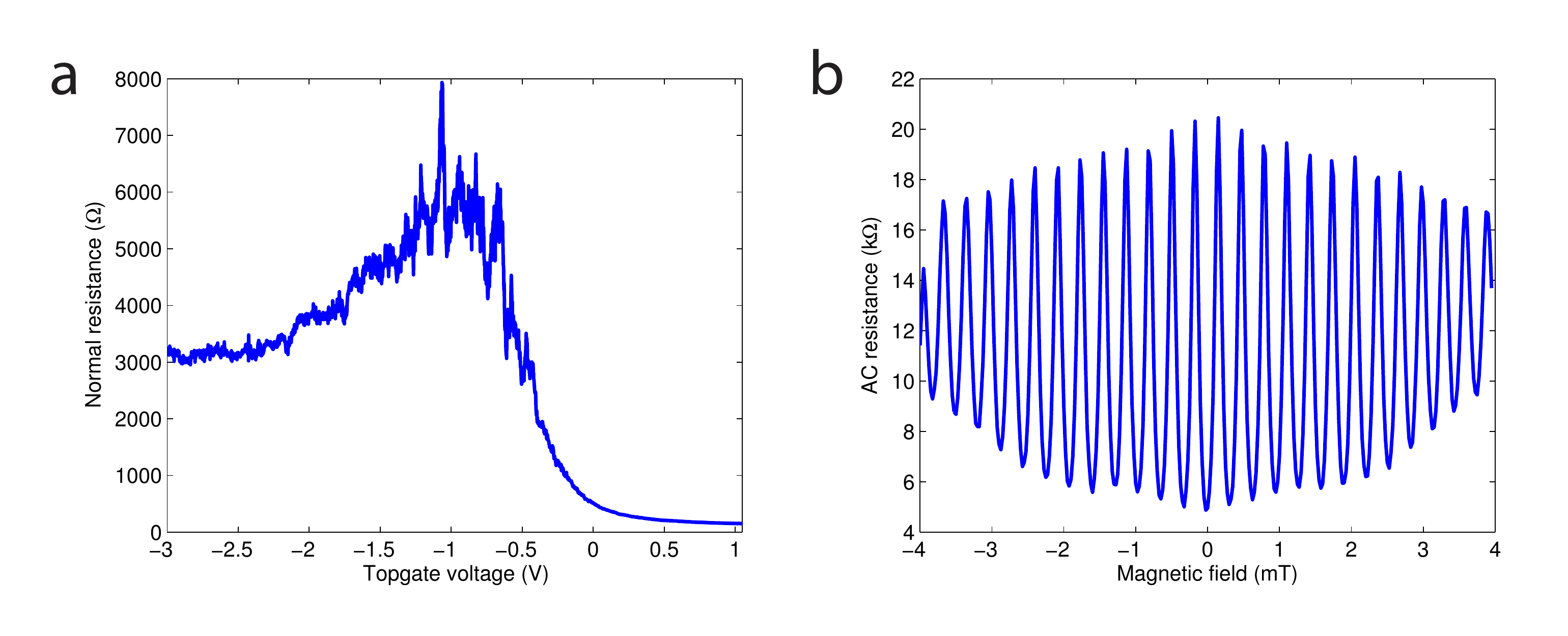}
\par\end{centering}

Supplementary Figure 6: Additional data for the topological junction
discussed in the main text, as the carrier density in the HgTe is
depleted even further. a) As a function of the topgate voltage the
normal AC resistance peaks near 6-8 k$\Omega$, consistent with the
QSH effect in the presence of several additional bulk modes. b) The
junction's AC resistance as a function of magnetic field, measured
with the topgate voltage at $V_{G}=-0.7$ V and with no DC current
bias. Even though the resistance minima are far from 0 $\Omega$,
the resistance oscillates with a period corresponding to $\Phi_{0}=h/2e$.
This periodic behavior is consistent with the superconducting interference
observed at higher densities, and suggests that supercurrent transport
persists well into the QSH regime. 
\end{figure}
\begin{figure}[p]
\begin{centering}
\includegraphics[scale=0.4]{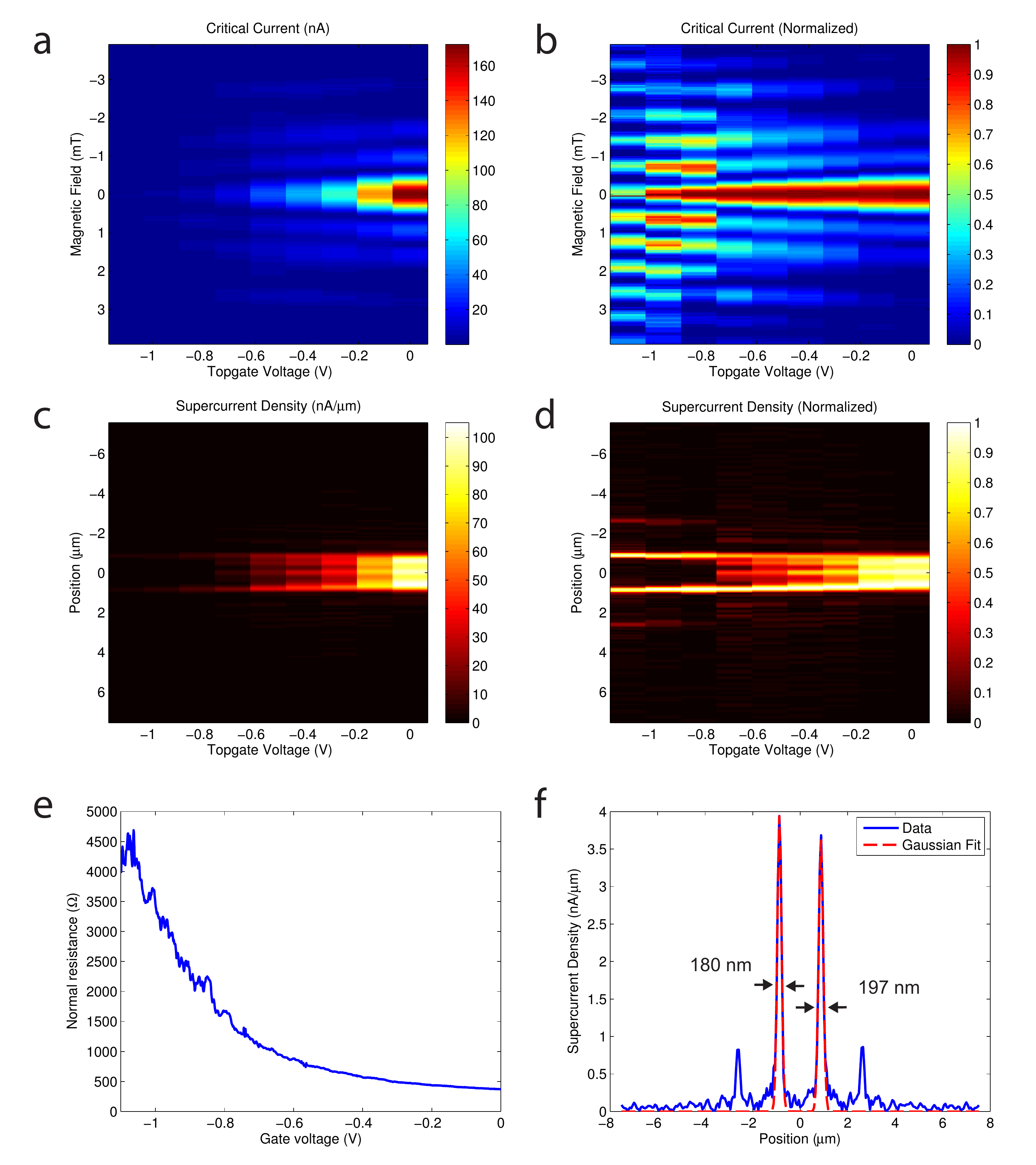}
\par\end{centering}

Supplementary Figure 7: Data from a Josephson junction fabricated
using Wafer I, the 7.5 nm quantum well. This junction was identical
to the one presented in the main text, except that the width of the
mesa is 2 microns. a) A map of the critical current envelope as a
function of topgate voltage shows that this device has a magnetic
field period of 0.68 mT, consistent with the overall area of the device.
b) After normalization the interference patterns show the evolution
of this device into the QSH regime. The decay of the interference
envelope over roughly 4 mT in the QSH regime is determined by the
widths of the edge channels. c, d) The supercurrent density shows
the confinement of supercurrent to edge modes as the bulk density
is depleted. e) The normal resistance of the junction as a function
of the topgate voltage. f) Edge widths extracted from the supercurrent
density at the farthest negative gate voltage (-1.1 V).
\end{figure}
\begin{figure}[p]
\begin{centering}
\includegraphics[scale=0.4]{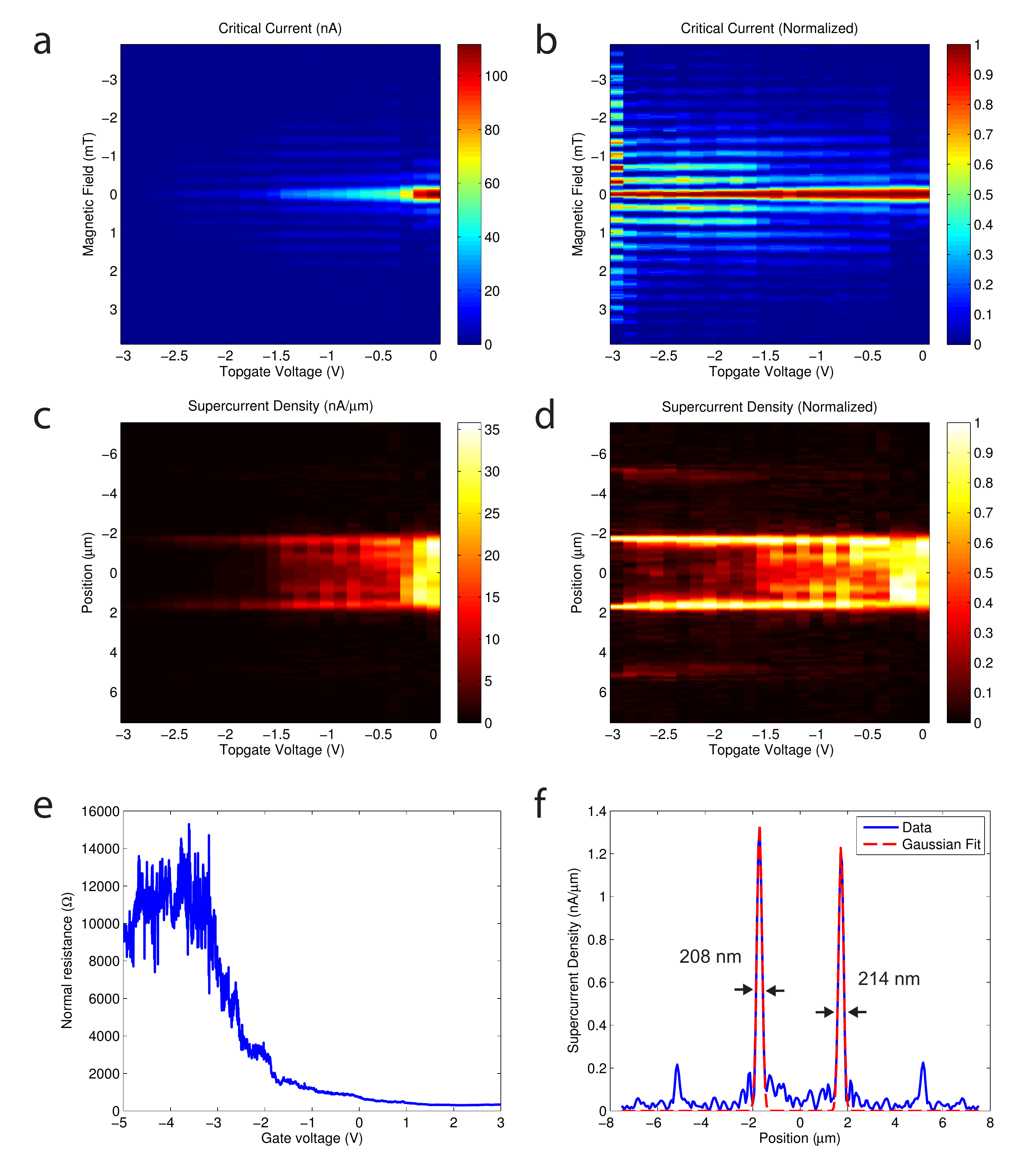}
\par\end{centering}

Supplementary Figure 8: Data from a Josephson junction fabricated
using Wafer I, the 7.5 nm quantum well. This junction was identical
to the one presented in the main text, except that the topgate was
200 nm and centered between the contacts. a,b) Consistent with other
topological devices, the critical current envelope tranforms from
a single-slit to a sinusoidal pattern as the density is decreased.
The decay of the interference lobes is over roughly 4 mT at the most
negative gate voltage. c, d) The supercurrent density shows the confinement
of supercurrent to edge modes as the bulk density is depleted. e)
The normal resistance of the device, extending beyond the $~4,000$
ohms where the smallest supercurrents were observed. The resistance
approaches the expected value for transport through two ballistic
QSH edges. f) Edge widths extracted from the supercurrent density
at the farthest negative gate voltage.
\end{figure}
}
\end{document}